\documentclass[twocolumn,pra,aps,showpacs,superscriptaddress]{revtex4-1}
\usepackage[english]{babel}
\usepackage[utf8]{inputenc}
\usepackage{graphicx,epsfig,placeins}
\usepackage{url,psfrag,graphicx}
\usepackage{dcolumn}
\usepackage{amsmath,amssymb,amsthm}
\usepackage{bm}
\usepackage{pstricks}
\usepackage{hyperref}
\usepackage{epsfig}
\usepackage{xcolor}

\usepackage[normalem]{ulem}

\def\<{\left<}
\def\>{\right>}
\def\ket|#1>{\left|#1\right>}
\def\bra<#1|{\left<#1\right|}

\def\elem<#1|#2|#3>{\left<#1\right|#2\left|#3\right>}
\def\({\left(}
\def\){\right)}
\def\[{\left[}
\def\]{\right]}

\def\H{{\mathcal H}}
\def\beq{\begin{equation}}
\def\eeq{\end{equation}}

\begin{document}

\title[Short Title]{Nanowire reconstruction under external magnetic
  fields}

\author{Eva M. Fernández}
\affiliation{Departamento de Física Fundamental, Universidad Nacional de
  Educación a Distancia (UNED), Madrid (Spain).}

\author{Silvia N. Santalla}
\affiliation{Departamento de Física \& GISC, Universidad Carlos III de Madrid,
  Leganés (Spain).}

\author{José E. Alvarellos}
\affiliation{Departamento de Física Fundamental, Universidad Nacional de
  Educación a Distancia (UNED), Madrid (Spain).}

\author{Javier Rodríguez-Laguna}
\affiliation{Departamento de Física Fundamental, Universidad Nacional de
  Educación a Distancia (UNED), Madrid (Spain).}

\date{December 2, 2020}

\begin{abstract}
We consider the different structures that a magnetic nanowire 
adsorbed on a surface may adopt under the influence of external 
magnetic or electric fields. 
First, we propose a theoretical framework based on an Ising-like 
extension of the 1D Frenkel-Kontorova model, which is analysed 
in detail using the transfer matrix formalism, determining a
rich phase diagram displaying structural reconstructions at 
finite fields and an antiferromagnetic-paramagnetic phase 
transition of second order. 
Our conclusions are validated using ab initio calculations with 
density functional theory, paving the way for the search of 
actual materials where this complex phenomenon can be observed 
in the laboratory.
\end{abstract}

\maketitle


\section{Introduction}

Surface atoms can behave in a very different way from their bulk
counterparts \cite{Oura.03}.  Their reduced coordination number
usually manifests itself in a change in the effective lattice
parameter, which induces stresses along the surface which can be
relaxed through a {\em surface reconstruction}, i.e. a full change of
symmetry of the surface structure, creating very interesting patterns.
Naturally, these reconstructions are also usual in the case of
heteroepitaxial systems, where film and substrate atoms belong to
different species \cite{Brune.94,Krzyzewski.01,Raghani.14}.  Moreover,
the same phenomenon can be considered in {\em nanowires}, quasi-1D
atomic structures, adsorbed on surfaces
\cite{Makita.03,Noguera.13,Yu.16,Lazarev.18}.  In any case, the
differences in energy of the different atomic configurations can be
quite small. Thus, predicting the configuration of minimum energy for
a homo- or heteroepitaxial system is a complex computational problem,
even when the interactions between the film and bulk atoms are known
\cite{Oura.03,Pushpa.09}.

Standard approaches include {\em ab initio} calculations such as
density functional theory (DFT), such as the studies of nanowires of
transition metals presented in \cite{Sargolzaei,Tung,Zarechnaya}. The
large computational cost demanded by large scale DFT simulations
suggests complementing them with effective statistical mechanics
approaches, such as the Frenkel-Kontorova (FK) model
\cite{Frenkel.38,Frenkel.39}, which has been extensively used to
describe the dynamics of adsorbate layers on a rigid substrate
\cite{Braun.04}.  In its original formulation, the FK model
represented the film of adsorbate atoms as point-like masses joined
with springs (i.e., nearest neighbor interactions), sitting on a rigid
periodic potential energy representing the substrate.  When the
natural length of the springs and the substrate periodicity differ,
the equilibrium configurations can become very rich
\cite{Noguera.13,Mansfield.90,Braun.04}.  Many extensions of the FK
model have been proposed, such as allowing for more realistic film
potentials, tiny vertical displacements \cite{Laguna.05} or even
quantum behavior of the film atoms \cite{Hu.00}.  Interestingly, FK
can be complemented with small-scale DFT calculations in order to fix
the form of the interaction, resulting in accurate predictions both
for the equilibrium and the kinetic effects
\cite{Pushpa.03,Pushpa.09}.

In this work we explore the possibility of obtaining different
nanowire structures when external fields, either electric or magnetic,
are applied.  If the energetic differences are tiny, external fields
can change notably the electronic configuration, effectively
preventing certain bonds or enhancing others, thus giving rise to
subtle changes in the surface lattice parameters.  Indeed, both bulk
magnetoelastic lattice distortions \cite{Barbara.77,Toft_Petersen.18}
and spin-phonon interactions \cite{Mattuck.60,Tschernyshyov.11} have
attracted considerable interest.  Moreover, examples of magnetization
mediated surface reconstructions have been reported
\cite{Teraoka.92,Gallego.05}, and the complementary concept of {\em
  magnetic reconstruction}, where the surface spins present a
different symmetry from the bulk, has also been discussed in the
literature \cite{Tang.93,Rettori.95,Maccio.95,Zhang.18}.

In this paper we propose a theoretical framework, which we term {\em
  Ising-Frenkel-Kontorova} (IFK) model, an extension of the 1D FK
model where the film atoms possess an Ising-like spin that can point
either up or down. When two neighboring film atoms have the same spin
their interaction is different from the case in which they have
opposite spins. An external magnetic field, then, can polarize the
spins, forcing them to adopt a parallel spin configuration and,
therefore, to change their equilibrium configuration. As temperature
increases the system undergoes a second-order phase transition from an
anti-ferromagnetic to a paramagnetic configuration, which we
characterize using the transfer operator formalism and finite-size
scaling of the magnetic susceptibility. The results of the statistical
mechanics approach are then tested using {\em ab initio} calculations
of chains of H and Fe, showing that the physical predictions are
qualitatively consistent.

This article is organized as follows. In Sec.~\ref{sec:ifk} we
describe the IFK model in detail, along with numerical results about
the phase diagram. The {\em ab initio} calculations are carried out in
Sec.~\ref{sec:dft}. A unified physical picture, combining the results
from the two different approaches, can be found in
Sec.\ref{sec:picture}. The article ends with a presentation of our
conclusions and our proposals for further work.


\section{The Ising-Frenkel-Kontorova model}
\label{sec:ifk}

Let us consider a simple extension of the 1D FK model, that we have
termed Ising-Frenkel-Kontorova (IFK), which consists of adding an
Ising spin variable, $+$ or $-$, to each film atom, representing its
spin polarization along a certain {\em easy axis}.  We will only
consider {\em coherent} films, where the number of film and substrate
atoms is the same, and each film atom is always in correspondence
with a substrate atom.

Let $r_i$ be the position of the $i$-th atom, and $s_i$ be its spin
polarization. The total Hamiltonian of the model for $N$ atoms is:
\begin{equation}
  \H = \sum_{i=1}^N \( V_s(r_i) -H s_i \)
  + \sum_{i=1}^{N-1} V_f\(|r_i-r_{i+1}|,s_is_{i+1}\),
\label{eq:ifk}
\end{equation}
where $V_s(r_i)$ stands for the (periodic and rigid) substrate
potential felt by each film atom, while $V_f(d,s_is_{i+1})$ represents
the atom-atom film interaction, which depends on their distance and
their relative polarization: if the two spins are parallel, the
interaction potential is $V_f(d,+1)$, and if they are anti-parallel,
it is $V_f(d,-1)$. Moreover, $H$ represents the external magnetic
field along the chosen axis. The Hamiltonian \eqref{eq:ifk} should be
interpreted as possessing open boundary conditions. When we minimize
that Hamiltonian we obtain a semi-classical configuration: positions
plus spin polarization of all atoms.

Notice that neighboring atoms can interact through two different
potential energy functions: a {\em ferro} (F) potential,
$V_F(d)=V_f(d,+1)$ or an {\em anti-ferro} (AF) one,
$V_{AF}(d)=V_f(d,-1)$. These two potentials can have different
equilibrium distances, $a_F$ and $a_{AF}$. Indeed, in some cases one
of them (typically, the ferro potential) may not present a minimum at
any distance, and $a_F$ cannot be defined.

\begin{figure}
  \includegraphics[width=8cm]{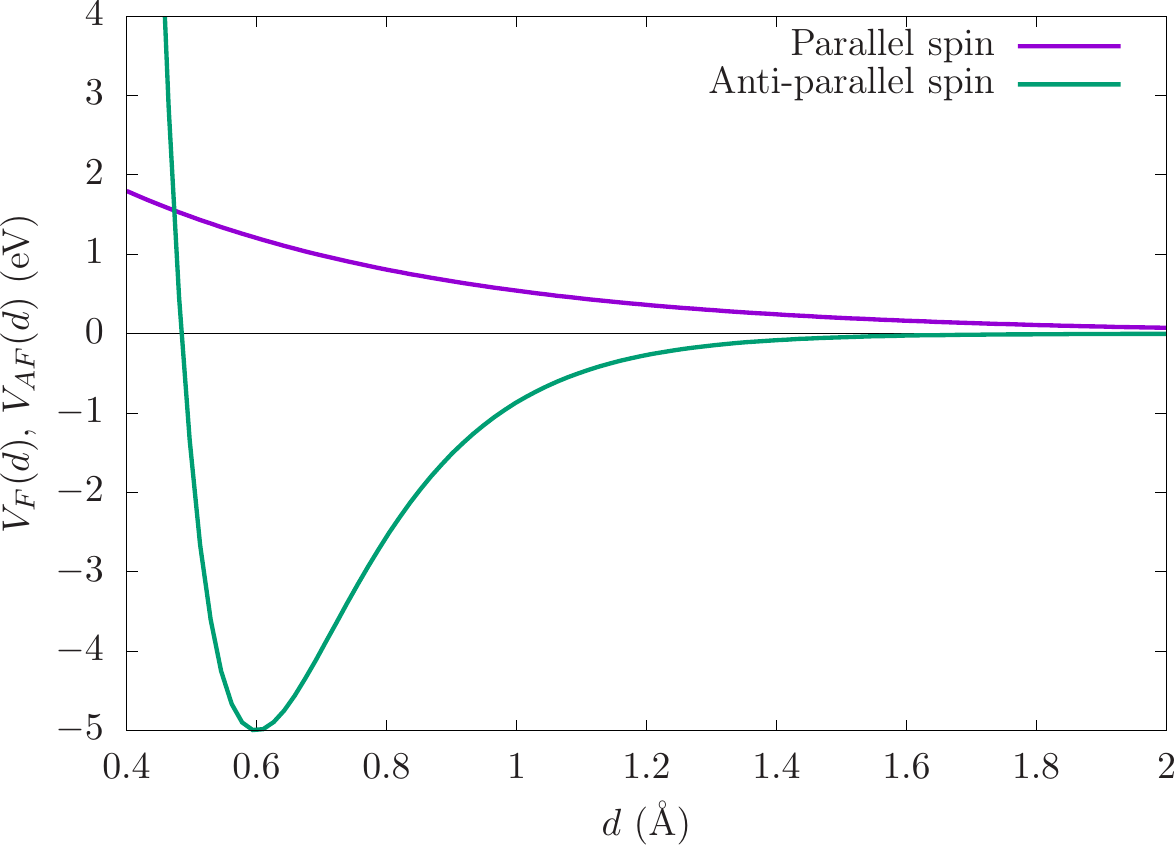}
  \caption{Atom-atom film potentials used in our calculations, 
    both in the ferro (parallel spins) and antiferro 
    (anti-parallel spins) configurations, see Eq. 
    \eqref{eq:potentials}.}
  \label{fig:illust}
\end{figure}

Let us particularize for the case shown in Fig.~\ref{fig:illust},
where we can see that $V_F(d)$ does not present a minimum, while
$V_{AF}(d)$ does, and let us assume that $a_{AF}\neq a_s$ (the lattice
parameter of the substrate). Let us also assume that the lowest
energy of the ferro potential exceeds the value for the antiferro
case, as it is usually the case. In absence of an external field,
there will be a {\em misfit} between the substrate and the film
lattice parameters and, if the substrate potential is small enough,
the film atoms can {\em reconstruct}.

Yet, when an external magnetic field is applied, at a certain moment
the ferromagnetic configuration will be preferred energetically.
Then, the advantage of reconstruction is lost, and, if the film
remains coherent it will {\em wet} the substrate, i.e. it will copy
its structure.

We will choose the following expressions for the three 
potential energy interactions:
\begin{align}
&  V_s(r) = V_{s,0} \ \cos(2\pi r/a_s),
\nonumber \\
&  V_{AF}(d)  = 
V_{AF,0} \ \Big(1 - e^{-b_{AF}(d-a_{AF})} \Big)^2 - V_{AF,0},
\nonumber \\
&  V_F(d)  = V_{F,0}\ \exp(-b_F d),
  \label{eq:potentials}
\end{align}
i.e. a sinusoidal form for the film-substrate potential, a Morse form
for the AF film potential and an exponential decay for the F film
potential.

In our calculations we will employ $k_B=1$, measuring temperatures in
energy units, which we choose to be eV.  Also, the magnetic field $H$
will be measured in energy units, by making the Bohr magneton
$\mu_B=1$.  Thus, $H=1$\,eV corresponds approximately to $2\cdot
10^4$\,T when the spin values are $S_z=\pm \hbar/2$, which we
normalize to be $s=\pm 1$.  For the sake of concreteness, we will use
the following parameters for the effective potentials: $a_s=1$\,\AA,
$V_{s,0}=2$\,eV, $V_{F,0}=4$\,eV, $b_F=2$\,\AA${}^{-1}$,
$V_{AF,0}=5$\,eV, $b_{AF}=6$\,\AA$^{-1}$, $a_{AF}=0.6$\,\AA, which
constitute a reasonable choice suggested by the {\em ab initio}
calculations for H chains as provided in Sec.~\ref{sec:dft}.
Fig.~\ref{fig:illust} shows the curves for $V_{AF}(d)$ and $V_{F}(d)$
using these values.


\subsection{Transfer operator approach}

The physical properties of the system described by 
Hamiltonian \eqref{eq:ifk} in equilibrium at 
temperature $T=\beta^{-1}$ are determined by the 
partition function:
\beq
Z=\sum_{\{r_i,s_i\}} \exp\[-\beta \H(\{r_i,s_i\}) \].
\label{eq:z}
\eeq

Since the system is one-dimensional, we can write this 
partition function as a trace over a product of transfer 
matrices \cite{Baxter.82}. 
It is convenient to introduce new notation to simplify 
our expressions. 
Let $x_i=\{r_i,s_i\}$ denote the multi-index which 
combines the position and the spin of the $i$-th atom.  
Then, the IFK Hamiltonian, Eq. \eqref{eq:ifk} can be 
written as a sum of a one-body and a two-body terms
\beq
\H = \sum_{i=1}^N \H^{(1)}(x_i) + \sum_{i=1}^{N-1}\H^{(2)}(x_i,x_{i+1}),
\label{eq:Hdecomp}
\eeq
with $\H^{(1)}(x_i)=V_s(r_i)-Hs_i$ and
$\H^{(2)}(x_i,x_{i+1})=V_f(|r_i-r_{i+1}|,s_is_{i+1})$. 
Let us consider $x_i$ to be restricted to take only a 
value from a finite set with $\ell$ elements. 
Then, we can define
\begin{align}
  V_{x_i} &\equiv \exp(-\beta \H^{(1)}(x_i)), 
 \nonumber \\  \nonumber \\
  T_{x_i,x_{i+1}} &\equiv \exp(-\beta \H^{(2)}(x_i,x_{i+1})), 
\nonumber \\  \nonumber \\
  M_{x_i,x_{i+1}} &\equiv T_{x_i,x_{i+1}} V_{x_{i+1}},
\label{eq:transfer}
\end{align}
leaving the dependence on the parameters ($\beta$, $H$, etc.)
implied. 
Now, $V$ is a vector with $\ell$ components, $T$ and $M$ are
matrices with dimension $\ell\times\ell$. 
We can then write:
\begin{align*}
  Z &=
\sum_{\{x_i\}} V_{x_1} T_{x_1,x_2} V_{x_2} T_{x_2,x_3} V_{x_3}
  \cdots T_{x_{N-1},x_N} V_{x_N} 
\nonumber\\
  &=\sum_{\{x_i\}} V_{x_1} M_{x_1,x_2} M_{x_2,x_3} \cdots M_{x_{N-1},x_N}
\end{align*}
and taking into account that all $M$ matrices are equal 
(which need not be the case in a more general setting)
we have
\begin{align}
  Z &= \sum_{x_1,x_N} V_{x_1} \ (M)^{N-1}_{x_1,x_N} 
\nonumber \\
  &=V^T \ (M)^{N-1} \ S,
\label{eq:product}
\end{align}
where $S=(1,\cdots,1)^T$. 
The numerical evaluation of expression \eqref{eq:product} 
is a standard problem in statistical mechanics, which 
may be carried out through the spectral decomposition 
of $M$ \cite{Baxter.82}.

Expectation values are obtained by inserting appropriate 
operators in the matrix product. 
Let us consider $\ell$ component vectors $R_{x_i}$
and $S_{x_i}$, which measure the expectation value of 
the position and spin of the $i$-th atom: 
$R_{x_i}=r_i$, $S_{x_i}=s_i$. 
Then,
\begin{align}
\<r_i\>&= {1\over Z} \sum_{\{x_i\}} V_{x_1} M_{x_1,x_2}
\cdots  R_{x_i}M_{x_i,x_{i+1}} \cdots M_{x_{N-1},x_N}
\nonumber  \\  \nonumber  \\
\<s_i\>&= {1\over Z} \sum_{\{x_i\}} V_{x_1} M_{x_1,x_2}
\cdots  S_{x_i}M_{x_i,x_{i+1}} \cdots M_{x_{N-1},x_N}.
\label{eq:measures}
\end{align}
Moreover, the two-point correlators can be found in a 
similar way, inserting two operators, 
e. g. $R_{x_i} S_{x_i}$. 
The total magnetization $m\equiv \sum_i s_i$ can be 
obtained more succintly as
\begin{equation}
  m(\beta, H) = - {1\over\beta} {\partial\log Z\over \partial H}.
  \label{eq:magnZ}
\end{equation}

We would like to stress the similarity between expression
\eqref{eq:product} and a matrix product state (MPS) \cite{PGarcia.07},
with the different positions of the particles, $r_i$, playing the role
of the ancillary space, and the number $\ell$ of different positions
being the bond dimension. Physically, the bond dimension of an MPS
bounds the amount of information that we need to keep from the left
part of the chain in order to determine the probability for the
configuration of the right part. Thus, in our case this information
is represented by a continuous variable, allowing for a richer
behavior than in the case of the standard Ising model.


\subsection{Numerical Results}

The formalism presented in the previous section can be extended easily
to continuous values of $r_i$.  Yet, for practical calculations, it is
convenient to consider a suitable discretization.  A straightforward
strategy would be to consider a length $L$ sufficiently large to hold
the full chain of atoms, and to discretize it into small intervals of
length $\Delta x$, studying the limit $\Delta x\to 0$.  Taking spin
into account, this would give a matrix size $\ell=2L/\Delta x$.  In
practice, this leads to working with large matrices.

In this work we will only consider coherent films, with the 
same density as the substrate, and with only one film atom 
per unit cell. 
Thus, each $r_i \in [0,a_s]$, with $i = 1, \dots, \ell$,
and the discretization step must be taken as 
$\Delta x = a_s/(\ell-1)$. 
Thus, the dimensions of the matrices will always be
$2\ell\times 2\ell$.

\begin{figure}
  \includegraphics[width=9cm]{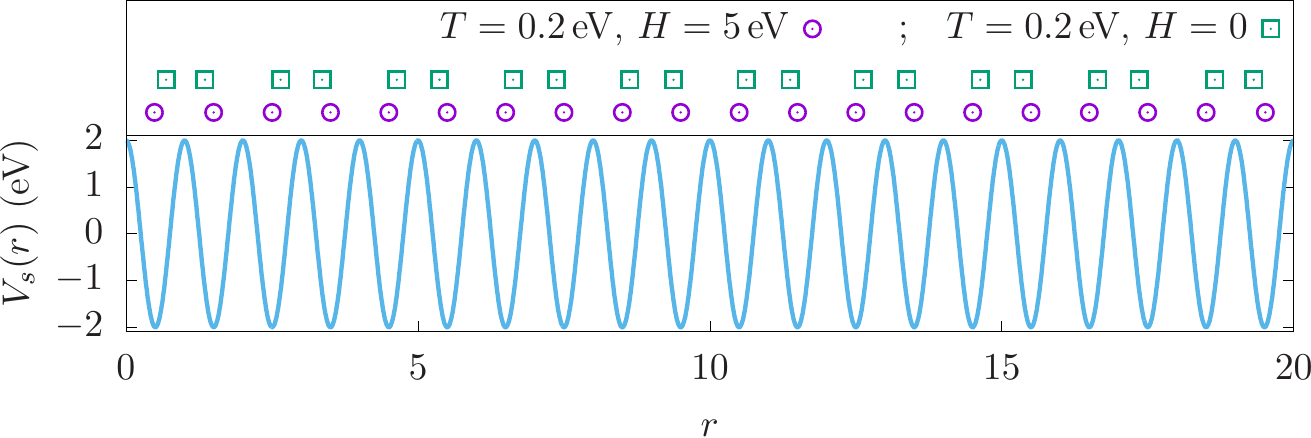}
  \caption{Average atom positions of the IFK model with $N=20$ using
    the parameters described in the text at very low temperature,
    $T=0.2$ eV, using zero magnetic field, $H=0$ eV and a large
    magnetic field, $H=5$ eV. Notice that the system dimerizes in
    absence of magnetic field. The blue line represents the
    film-substrate potential.}
  \label{fig:conf}
\end{figure}

\begin{figure}
  \includegraphics[width=8cm]{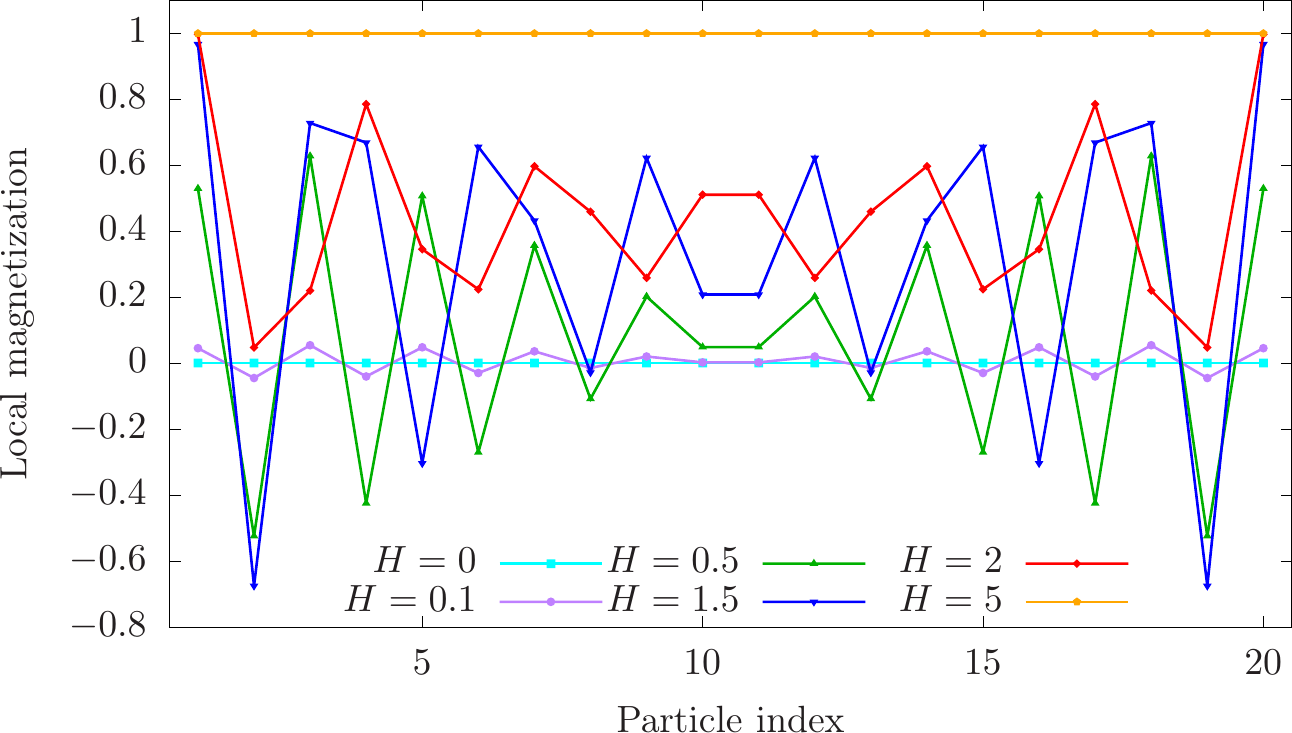}
  \includegraphics[width=8cm]{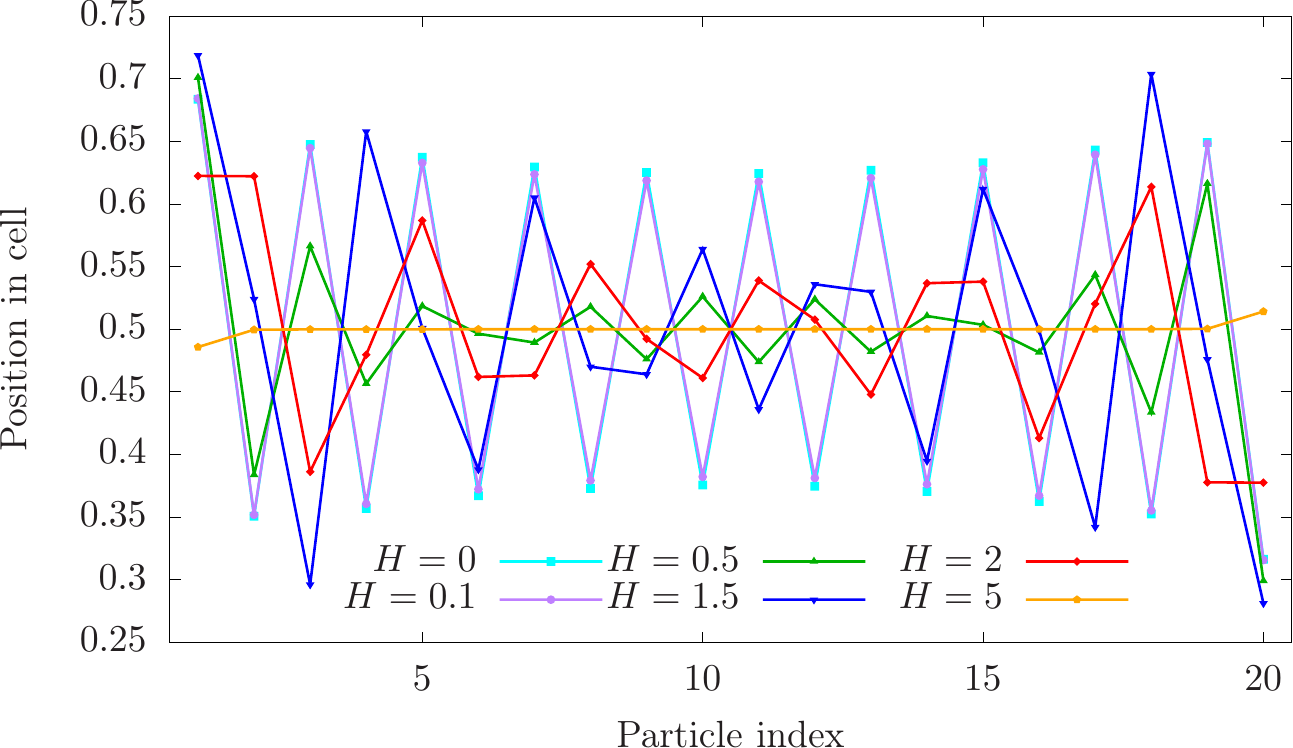}
  \caption{Top: average local magnetization of each atom in the open
    chain with $N=20$, for $T=0.2$\,eV and different values of the
    magnetic field (in eV). Bottom: average position of each film atom
    within the unit cell of the substrate, for the same values of
    temperature and magnetic field.}
  \label{fig:measures}
\end{figure}

We have computed exactly the partition function for an open chain,
obtaining the expected positions of the film atoms making use of
Eqs.~\eqref{eq:measures}. Fig.~\ref{fig:conf} shows these values at a
low temperature, $T=0.2$ eV, for both $H=0$ and $H=5$ eV.  We can see
that, for $H=0$ eV the system dimerizes, i.e.  presents an elementary
reconstruction, doubling its unit cell. Indeed, for $H=0$ the system
is antiferromagnetic, and becomes ferromagnetic, with almost all its
spins parallel when $H \neq 0$ (as can also be clearly seen in
Fig.~\ref{fig:measures}).

\begin{figure}
  \includegraphics[width=8cm]{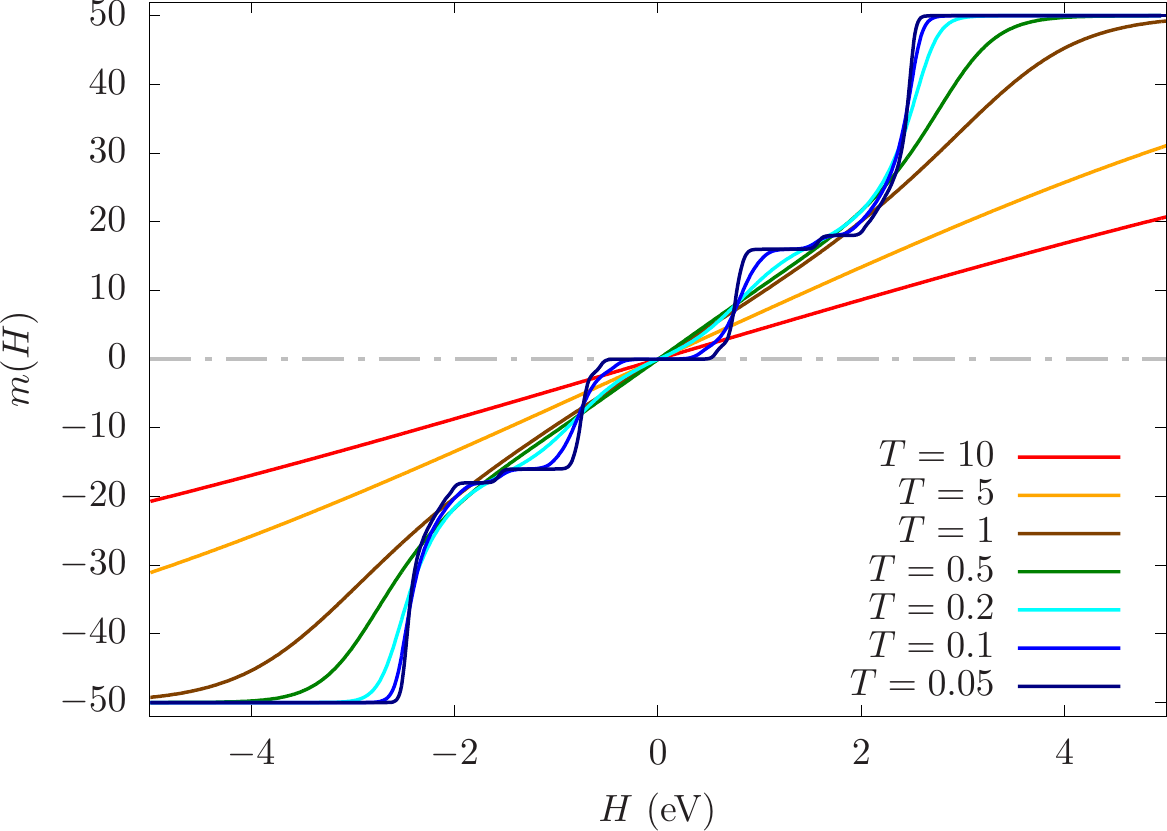}
  \caption{Magnetization curves $m(H)$ for the IFK model with $N=50$
    using the parameters discussed in the text, for several
    temperatures, ranging from $T=0.05$\,eV to $T=10$\,eV.}
  \label{fig:magn}
\end{figure}

The total magnetization curve, $m(H)$ corresponding to a system with
$N=50$ atoms is shown in Fig.~\ref{fig:magn} for a range of
temperatures spanning from $T=0.05$\,eV to $T=10$\,eV.  We can see
that, for high temperatures, the film is completely paramagnetic, with
a nearly constant slope for $m(H)$ even for $T=1$\,eV.  For
temperatures below $T=0.2$\,eV, we can observe a sharp increase in the
magnetization around $H_c \approx \pm 2$\,eV, which could correspond
to a paramagnetic-ferromagnetic transition. When $T$ is below
$0.1$\,eV there appears another sharp increase in $m(H)$ around $H_c
\approx \pm 1$\,eV, with a plateaux between them.  As a result, the
critical temperature must be around $T=0.1$\,eV.

Yet, the atomic and spin configurations of each film atom for low
temperatures can be rather complex, as we can see in
Fig.~\ref{fig:measures}. The top panel of this figure shows the
average magnetization of each atom, $\<s_i\>$, evaluated via
Eq.~\eqref{eq:measures}, for different values of the external magnetic
field, $H$ for an IFK model with $N=20$ atoms (instead of $N=50$, for
easier visualization). For $H \sim 0$, the average magnetization is
close to zero, increasing in amplitude near the borders, but keeping
an approximate anti-ferromagnetic pattern. The outer spins,
nonetheless, tends to be parallel to the external magnetic field, thus
explaining the increase in the average magnetization, but holding a
frustrated structure in the interior, because the number of atoms is
even. The local magnetization pattern, as we can see, is complex for
intermediate values of the magnetic field, becoming fully
ferromagnetic only for very large values of $H$.

The bottom panel of Fig.~\ref{fig:measures} shows the position of each
film atom within the unit cell of the substrate, with
$x=0.5$\,\AA\ denoting its center, always assuming $T=0.2$\,eV and the
same values as before for the external magnetic field.  For $H \sim
0$, we see the alternating pattern corresponding to the dimerized
reconstruction that we have shown in Fig.~\ref{fig:conf}.  Note that
in this case the position of each alternating atom shifts about 10\%
the size of the unit cell.  This pattern attenuates near the center as
the magnetic field increases, and for $H \sim 1.5$\,eV we can observe
a change of the deformation phase in the right extreme of the chain,
due to the fact that the rightmost extreme prefers to be polarized
along the direction of the external field.  For $H \sim 2$\,eV, the
whole pattern attenuates substantially, and for large magnetic fields
we can see that the film wets the substrate, copying its structure.
Notice that no frustrated structure appears in the interior. In any
case, we would like to stress that a chain with $N=21$ atoms will
yield the opposite behavior for both the magnetization and the
positions.

\begin{figure}
  \includegraphics[width=8cm]{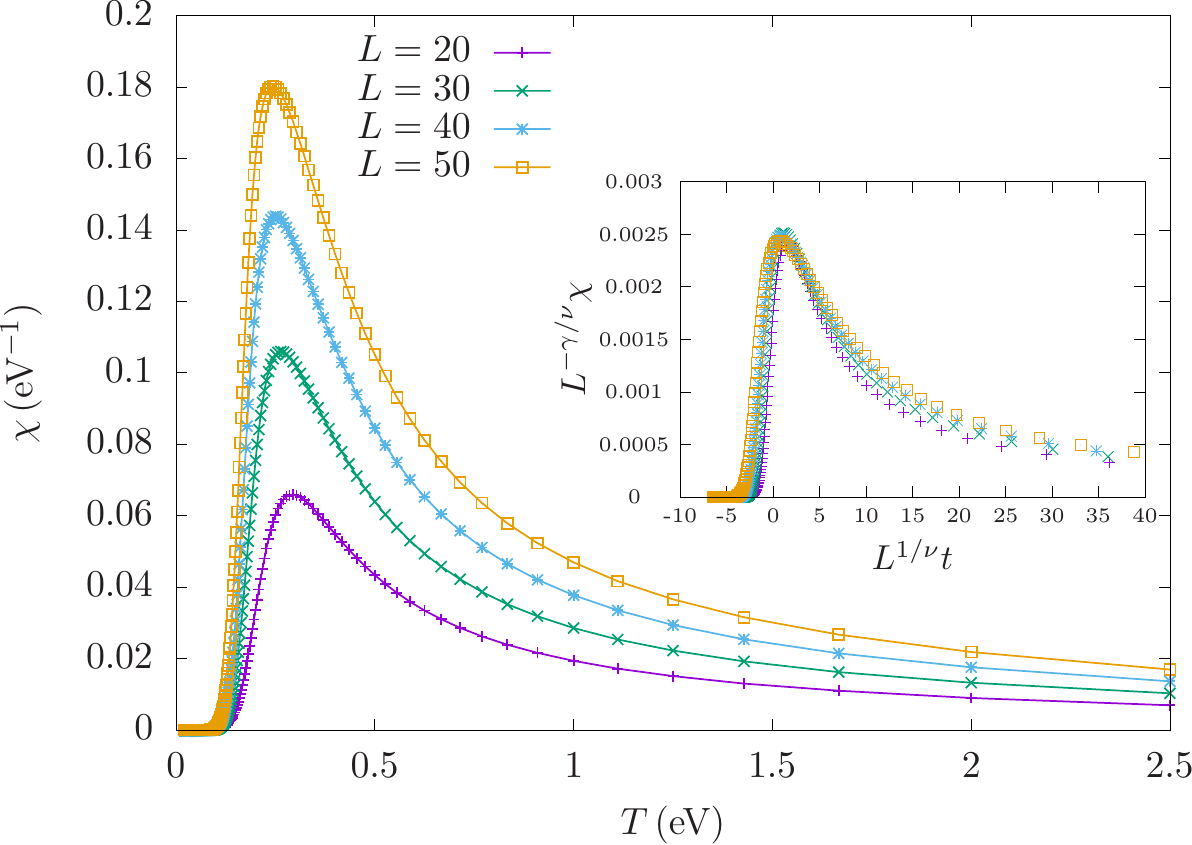}
  \caption{Magnetic susceptibility as a function of the temperatures
    for different system sizes. Inset: finite-size collapse of the
    susceptibility curves following Eq. \eqref{eq:finitesize}.}
  \label{fig:suscep}
\end{figure}

In order to obtain a full physical characterization of our 
system let us consider the behavior of the 
{\em magnetic susceptibility} $\chi$, defined as
\begin{equation}
  \chi=\left.{\partial m \over \partial H}\right|_{H=0},
  \label{eq:suscep}
\end{equation}
which is plotted in Fig.~\ref{fig:suscep}, for different system sizes.
We observe that, for low tempeatures, the susceptibility $\chi\approx
0$, which is consistent with the predicted antiferromagnetic (AF)
behavior.  At a finite temperature value $T \sim 0.15$\,eV we observe a
sharp rise, whose peak height depends on the system size $L$.
Furthermore, beyond the peak the susceptibility decays approximately
as $T^{-1}$, corresponding to the Curie law of paramagnetism.  This
behavior is consistent with a second-order phase transition from an
antiferromagnetic at low temperatures towards a paramagnetic behavior.
This type of phase transition is sometimes accompanied by structural
changes in Nature \cite{Zu.16,Liu.20}.

The nature of the phase transition and its critical 
exponents can be obtained through a finite-size scaling 
\cite{Newman.MCbook}, 
assuming that in the vicinity of the transition the 
susceptibility follows the law
\begin{equation}
  \chi(T)\approx L^{\gamma/\nu}F(L^{1/\nu}t),
  \label{eq:finitesize}
\end{equation}
where $L$ is the system size ($L\propto N$ in our case),
$t=(T-T_c)/T_c$ is the reduced temperature, $T_c$ is the critical
temperature, $\nu$ and $\gamma$ are the critical exponents associated
to the correlation length, $\xi\sim t^{-\nu}$, and the susceptibility,
$\chi\sim t^{-\gamma}$. The inset of Fig.~\ref{fig:suscep} shows that
an accurate collapse is obtained through $T_c\approx 0.11$ eV,
$\nu\approx 2.1$ and $\gamma\approx 2.31$.

Finding the mechanism behind these exponents is not an easy task.
Yet, we may conjecture that they may correspond to an Ising model with
long-range interactions \cite{Hiley.65,Glumac.89,Wragg.90,Cannas.95},
i.e. a Hamiltonian of the type
\begin{equation}
  {\cal H} = \sum_{i<j} {1\over |i-j|^\alpha} s_i s_j,
\end{equation}
where $\alpha$ is the decaying exponent for the coupling constants.
Indeed, the critical exponent for this model depends on $\alpha$. In
the range $\alpha \in (1,2)$ we find values for $\nu$ and $\gamma$
which are compatible with our results \cite{Wragg.90,Cannas.95}.

\bigskip

The rich phase diagram of the Frenkel-Kontorova model is determined by
two parameters: the lattice parameter misfit and the ratio between the
film and substrate potentials. In our IFK case, there are two
different film potentials, which give rise to two different possible
ratios. Yet, both are simultaneously changed when the amplitude of the
substrate potential is varied. Indeed, a change in $V_{s,0}$ can
induce a further phase transition, as we will describe below.

\begin{figure}
  \includegraphics[width=8cm]{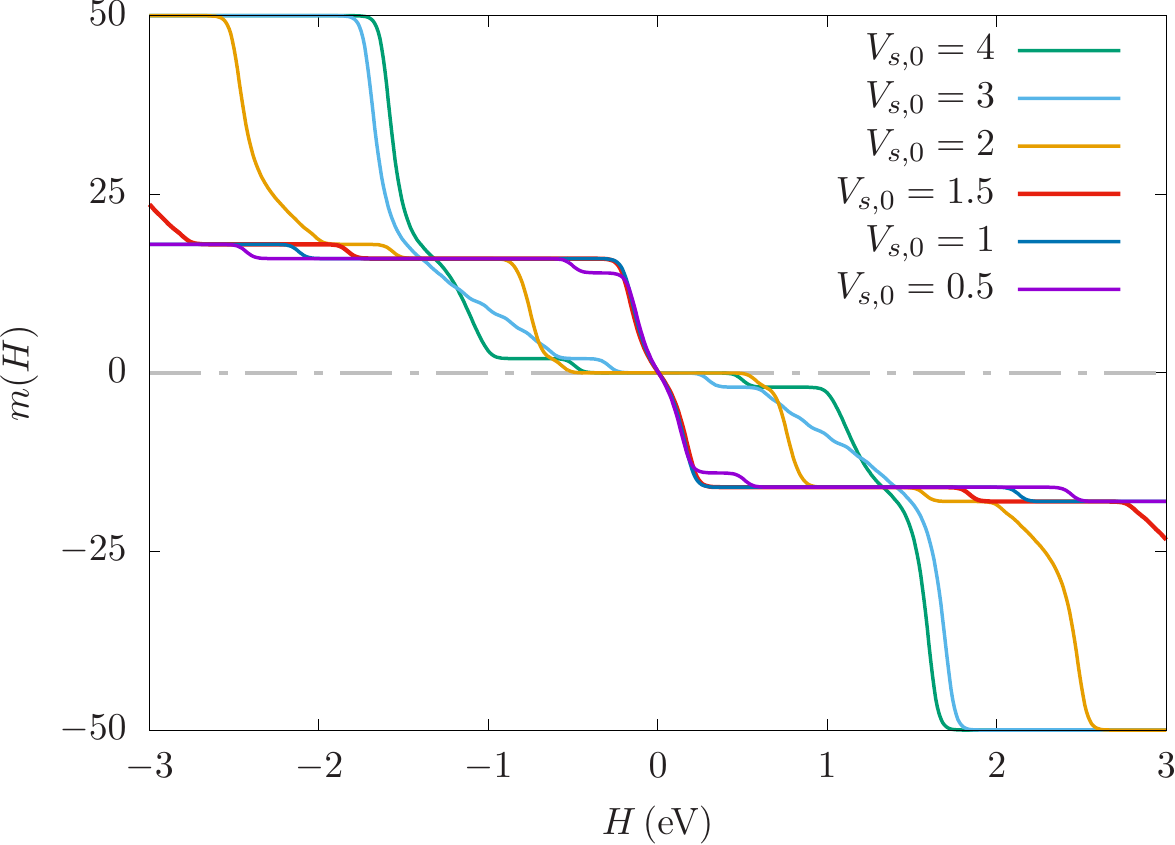}
  \includegraphics[width=8cm]{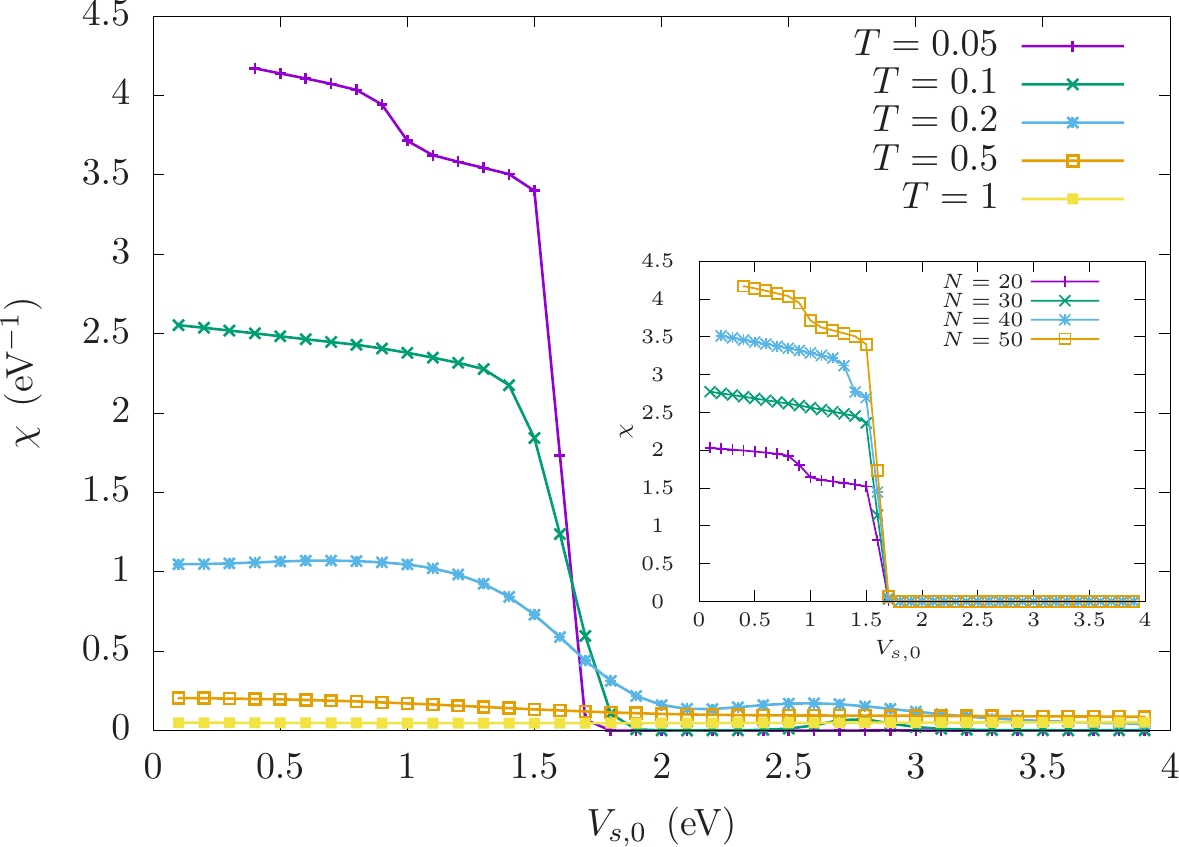}
  \caption{Top: magnetization as a function of the applied field,
    $m(H)$ using $N=50$ and $T=0.05$ for different values of $V_{s,0}$
    (in \AA). For $V_{s,0}$ above a critical value all magnetization
    curves present a finite plateau at $H=0$, which is a fingerprint of
    the antiferromagnetic phase. Bottom: susceptibility $\chi$ (at
    $H=0$) as a function of $V_{s,0}$, using $N=50$ and different
    values of $T$ (in eV). Notice that the susceptibility falls to
    zero above the aforementioned critical value unless the
    temperature is high enough. The inset shows the susceptibility as
    a function of $V_{s,0}$ for different system sizes $N$ using
    $T=0.05$.}
  \label{fig:vs}
\end{figure}

In Fig. \ref{fig:vs} (top) we can see the expected value of the
magnetization as a function of the applied magnetic field at very low
temperature ($T=0.05$) using $N=50$ atoms. All the parameters are the
same as in the previous calculations, except for the substrate
potential amplitude, $V_{s,0}$, which was varied around its original
value of 2\,eV. We should pay special attention to the vicinity of the
$H=0$ value, where we can see a finite plateau for high values of
$V_{s,0}$ and a finite slope for low values. This plateau is a
fingerprint of the antiferromagnetic phase, which can be seen to
disappear for weak substrate potentials (check Fig. \ref{fig:magn}). 

\begin{figure}
  \includegraphics[width=8cm]{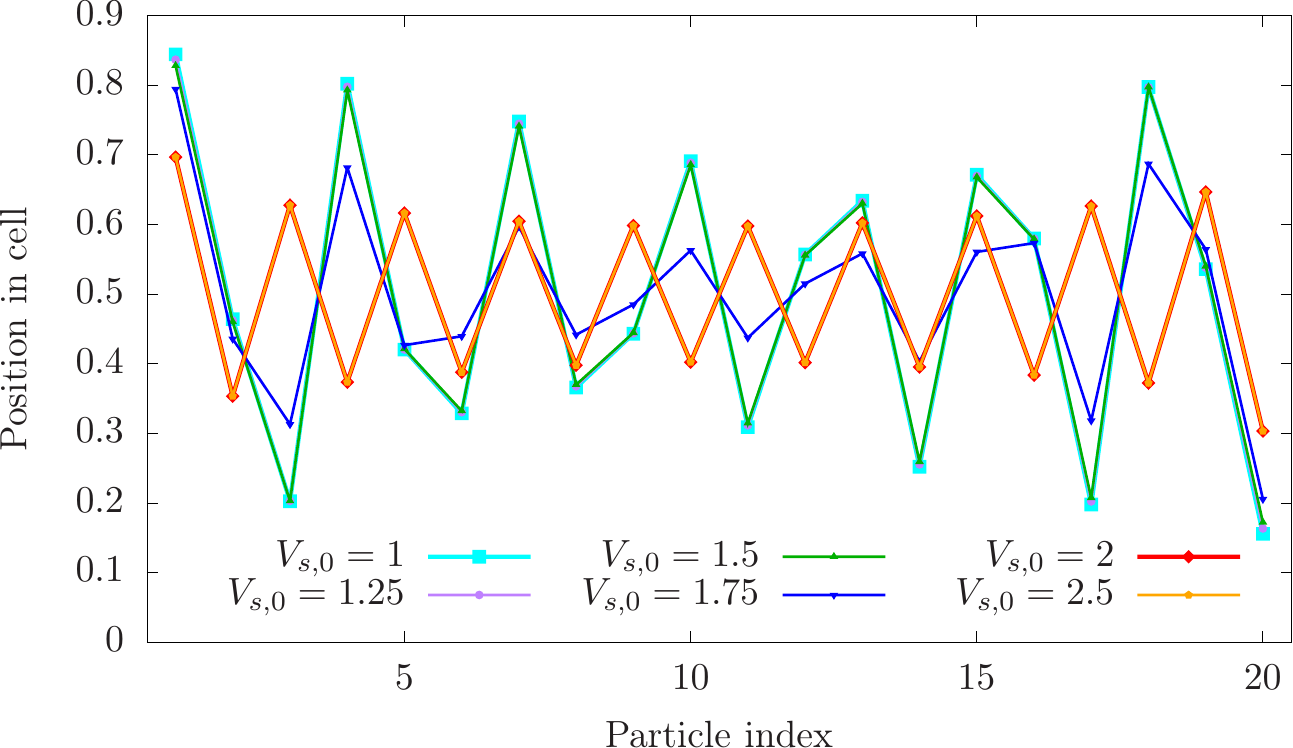}
  \caption{Atomic positions at $H=0$ for a chain with $N=20$ for
    easier visualization, using $T=0.05$ eV and different values of
    $V_{s,0}$. Notice that cases $V_{s,0}=1$, 1.25 and 1.5 eV collapse
    nearly exactly, because they correspond to the same structural
    phase, and the same can be claimed for $V_{s,0}=2$ and 2.5 eV,
    where the antiferromagnetic phase induces dimerization.}
  \label{fig:vs_pos}
\end{figure}

Yet, we can not simply claim that for low values of $V_{s,0}$ the
system reaches a paramagnetic phase. Indeed, the plateau exists for
all values of $V_{s,0}$, but it shifts away from $H=0$ when the
substrate potential is too weak. In the bottom panel of
Fig. \ref{fig:vs} we can see the dependence of the magnetic
susceptibility, $\chi$, (defined in Eq. \eqref{eq:suscep}) with
$V_{s,0}$ for different temperatures. We can observe a sudden drop for
low temperatures at a value $V_{s,0}\approx 1.75$ eV, while for high
temperatures the system becomes paramagnetic and $V_{s,0}$ becomes almost
irrelevant to determine $\chi$. The inset shows how the $T=0.05$ eV
curve changes when we choose different system sizes, and allows us to
claim that the jump in $\chi$ grows with $N$.

It is natural to ask whether this new transition has a visible
structural impact in the nanowire. We can see that this is indeed the
case in Fig. \ref{fig:vs_pos}, which shows the positions of the atoms
within the unit cell, in similarity to Fig. \ref{fig:measures}
(bottom), for different values of $V_{s,0}$, using always $N=20$ (for
easier visualization), $H=0$ and $T=0.05$ eV. Indeed, the
antiferromagnetic phase always corresponds to a nearly perfect
dimerization. Yet, for weak substrate potentials, $V_{s,0}<2$ eV we
observe a large deviation, with a period 3 modulation superimposed on
a smooth decreasing trend from the boundaries. This plot shows that
the magnetic structure interacts in a very non-trivial way with the
Frenkel-Kontorova degrees of freedom, giving rise to novel phenomena.

We would like to stress that our calculations always use open
boundaries, since they are the most natural setup for an atomic
nanowire. Moreover, the end atoms are always less attached to the
chain and are more susceptible to the action of an external
field. Moreover, also the parity of the number of atoms is
relevant. Indeed, if $N$ is even both end atoms can not align
simultaneously with the external field if the effective spin-spin
interaction is antiferromagnetic.


\section{Ab Initio Calculations}
\label{sec:dft}

In this section we show proof-of-principle ab initio calculations for
atomic chains (i.e. nanowires) performed with DFT. We have chosen two
different atomic species: on one hand, we have considered hydrogen
(H), because it gives rise to simple calculations. Moreover, we have
performed computations using iron (Fe), in order to compare with
previous {\em ab initio} studies of nanowires of transition metals
\cite{Sargolzaei,Tung,Zarechnaya}.

In both cases, we have built a chain of $N=8$ atoms with a total
length $L=a_sN$, where $a_s$ is the substrate lattice parameter, and
assuming periodic boundary conditions. Crucially, the expected value
of the total spin of the chain is fixed. For H, we have considered the
cases of $\<S_z\>=0$, $2\hbar$ and $4\hbar$ which, for $N=8$ atoms,
correspond to zero, half and full magnetizations. On the other hand,
for Fe we have considered $\<S_z\>=0$, $8\hbar$, $\dfrac{27}{2}\hbar$
and $16\hbar$, implying that the total magnetization is a fraction of
its maximal possible value: $0$, $0.25$, $0.42$ or $0.5$. In this way,
we will be able to characterize the behavior of the atom-atom film
interaction in absence of external magnetic field (zero magnetization)
or in the presence of external fields of given different strengths.
In order to simplify the calculations, the substrate potential is
absent from our calculations, except through the imposed substrate
lattice parameter.

\begin{figure}
  \includegraphics[width=8cm]{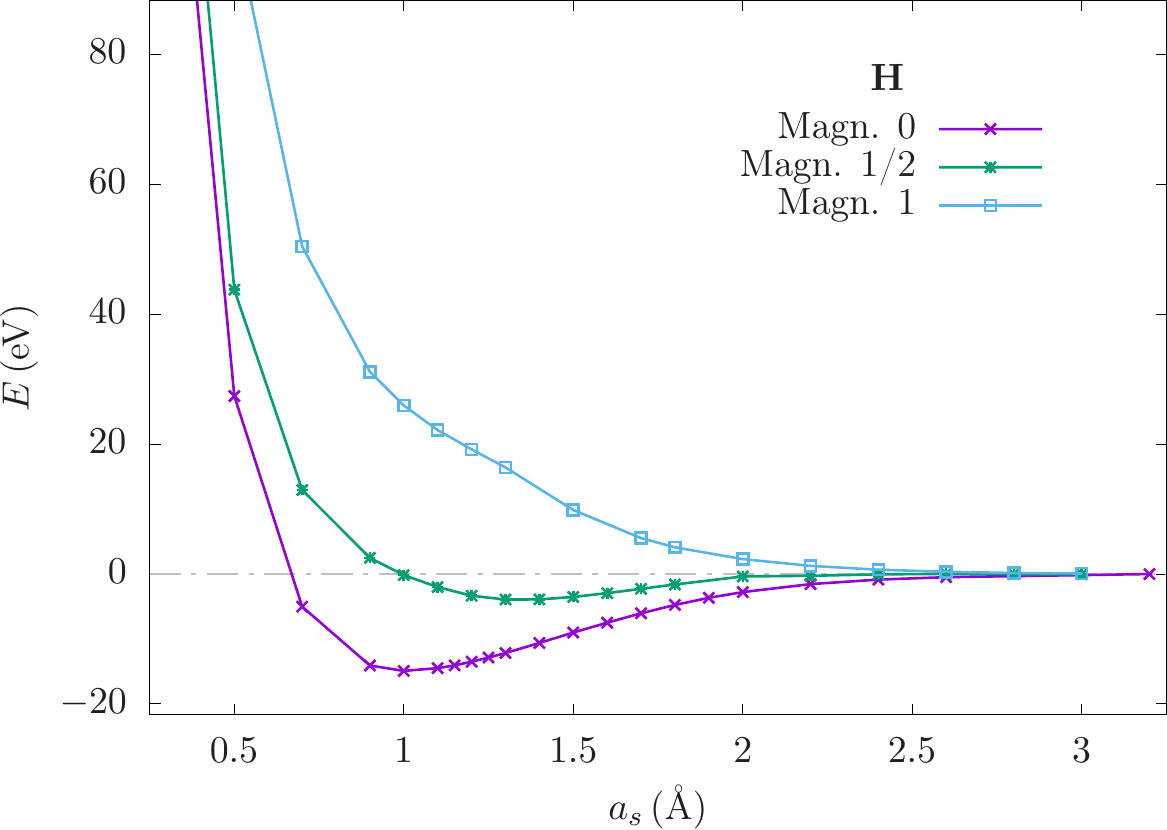}
  \includegraphics[width=8cm]{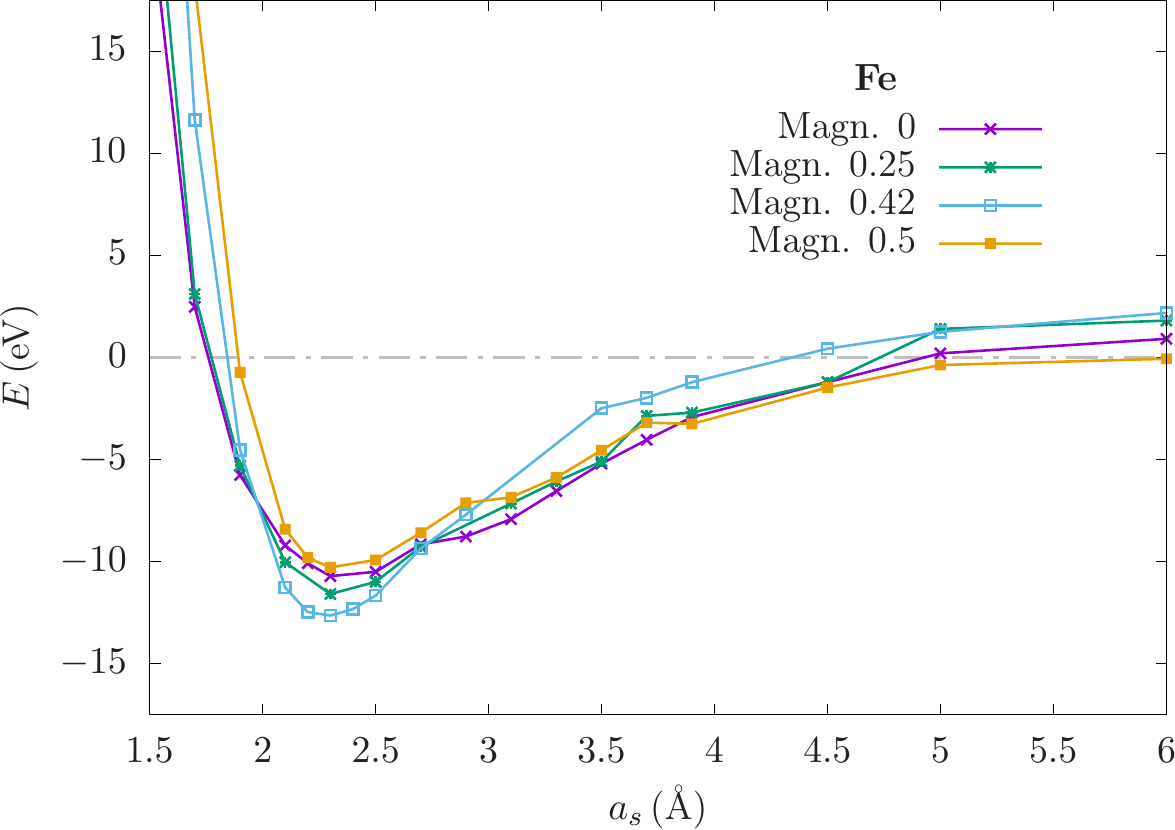}
\caption{DFT calculations of the total energy of a H (top) or Fe
  (bottom) chain of 8 atoms as a function of the lattice spacing,
  using periodic boundary conditions for several values of the
  magnetization fraction, i.e. $\<S_z\>$ normalized by the maximal
  possible magnetization, $n_e\hbar/2$.  In the case of Fe, notice the
  complex pattern of energy curves. Indeed, the optimal magnetization
  is non-zero for a range of values of $a_s$, while it may be zero for
  $a_s \in (2.8,3.5)$\,\AA.}
\label{fig:Edft}
\end{figure}

Electronic calculations were performed using the SIESTA code
\cite{Soler.02}, keeping fixed the chain structure during the
calculations, while the electronic part is relaxed. The exchange and
correlation potential was described using the Perdew, Burke, and
Ernzerhof (PBE) functional \cite{Perdew.96}. This functional was
already used in previous works on H$_2$ adsorption on single and
double aluminium clusters doped with vanadium or rhodium
\cite{Vanbuel.17,Vanbuel.19,Vanbuel.18,Jia.18}. The {\em core
  interactions} were accounted for by means of norm conserving scalar
relativistic pseudopotentials \cite{Troullier.91} in their fully
nonlocal form \cite{Kleinman.82}, generated from the atomic valence
configuration $1s^1$ for H and 4$s^2$3$d^6$ for Fe.  The core radii
for the $s$ orbital of H is $1.25$ a.u.  and for the $s$ and $d$
orbitals of Fe is 2.0 a.u.  The matrix elements of the self-consistent
potential were evaluated by integrating in an uniform grid.  The grid
fineness is controlled by the energy cutoff of the plane waves that
can be represented in it without aliasing (150 Ry in this work).
Flexible linear combinations of numerical pseudo-atomic orbitals (PAO)
are used as the basis set, allowing for multiple-$\zeta$ and
polarization orbitals.  To limit the range of PAOs, they were slightly
excited by a common energy shift (0.005 Ry in this work) and truncated
at the resulting radial node, leading to a maximum cutoff radii for
the $s$ orbitals of 6.05 a.u.  for H and 7.515 a.u. for Fe.  The chain
structure remains fixed during the calculations, while the electronic
part is relaxed.

As commented, in order to make proper comparisons wit our previous
results regarding the IFK model, and for the sake of saving
computational effort, we have considered two types of calculations
assuming fixed 1D chains (with 8 atoms of H or Fe), keeping constant
the expected value of the total spin of the chain, as discussed above.
Firstly, we have evaluated the total energy of the chains as a
function of the varying distance between adjacent atoms. Secondly, we
have assumed a fixed lattice parameter for the unit cell and evaluated
the energy of dimerized chains for distinct values of the dimerization
parameter.

In the first numerical experiment, we have calculated the total energy
of the chain as a function of the substrate lattice spacing, $a_s$,
assuming that the film copies the substrate, for all magnetizations
discussed previously.  The results are presented in
Fig.~\ref{fig:Edft}, where the top panel represents the results for
the H chains and the bottom panel those for the Fe ones.  Notice that
the energy zero has been set to the (lowest) energy obtained for the
largest value of $a_s$ for a better visualization.  The results are
obtained for the selected values of the magnetization fraction,
i.e. the expected value of the total spin of the chain, $\<S_z\>$,
divided by the maximal possible value, $n_e\hbar/2$, where $n_e$ is
the total number of electrons (8 for H and 64 for Fe).  We can observe
a similar behavior to the potentials between the film atoms in
Fig.~\ref{fig:illust}. For hydrogen, we see that in the absence of an
external magnetic field the system will choose the configuration with
zero magnetization, with an energy minimum around a value $a_s \approx
1$\,\AA.  As the magnetic field increases, the magnetic contribution
to the total energy will eventually favor the upper curves,
corresponding to higher total spin.  For Fe, on the other hand, for no
magnetic field the preferred magnetization is $\<S_z\> \neq 0$, and
the behavior of the energy curves are sufficiently different to
suggest that the presence of external magnetic fields will give rise
to different film interaction potentials.  The results for H (top
panel of Fig.~\ref{fig:Edft}) inspired our choice of values of the
physical parameters of the IFK model discussed in Sec.~\ref{sec:ifk}.

\begin{figure}
  \includegraphics[width=8cm]{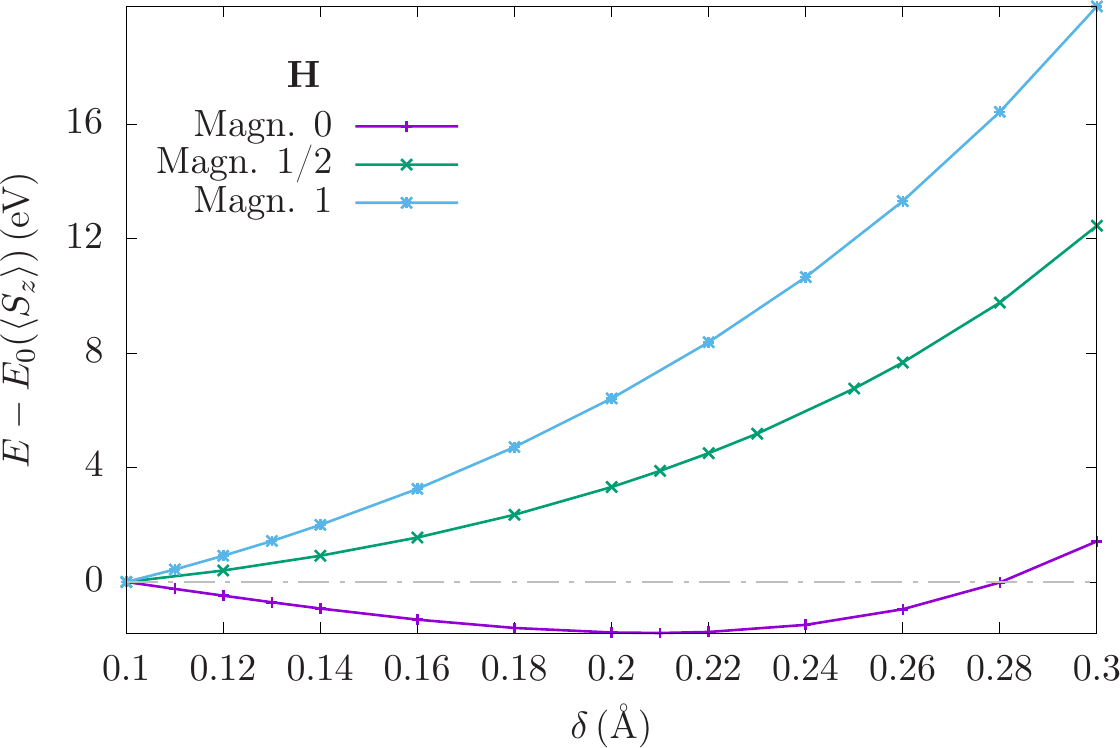}
  \includegraphics[width=8cm]{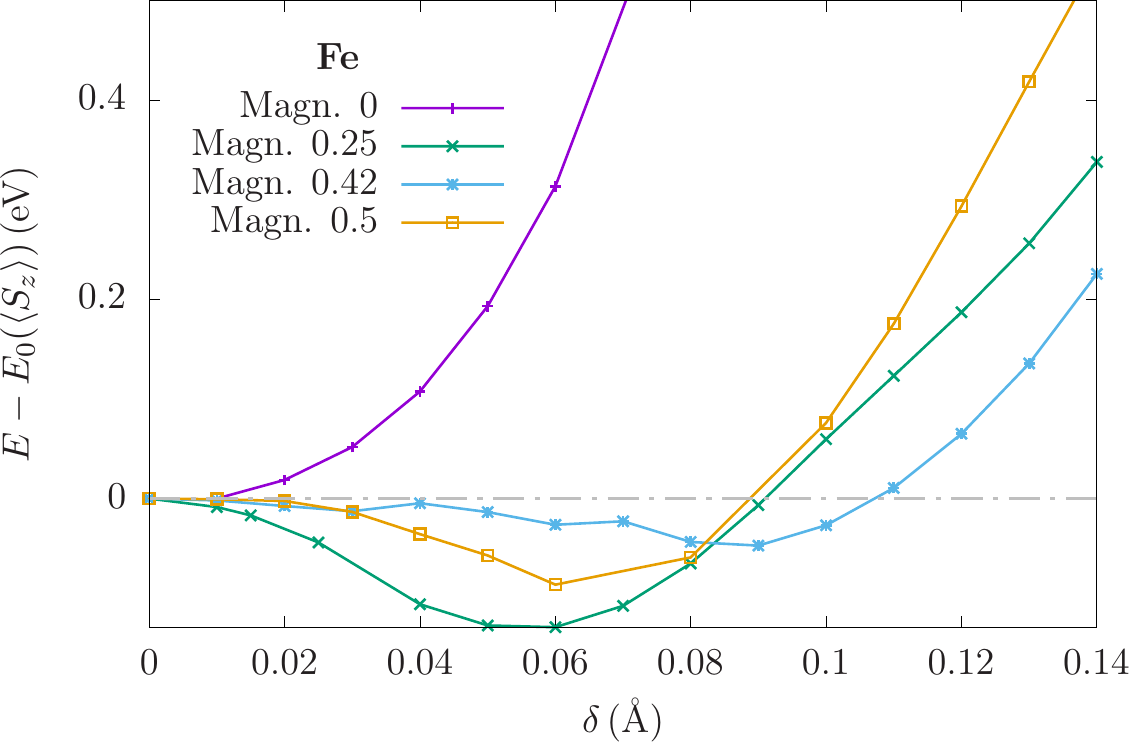}
  \caption{Energy of the same chains as in Fig.~\ref{fig:Edft} when
    allowed to dimerize, as a function of the dimerization parameter
    $\delta$, using $a_s=1.0$\,\AA\ for H (top) and $a_s=2.3$\,\AA\ for
    Fe (bottom), for the same values of the magnetization $\<S_z\>$,
    as in Fig.~\ref{fig:Edft}. The curves are vertically shifted a
    quantity $E_{\delta=0}(\<S_z\>)$ for a better comparison.}
  \label{fig:dimerdft}
\end{figure}

For the second computer experiment we fix the lattice parameter of
the substrate to the minimun obtained in the Fig.~\ref{fig:Edft}
($a_s=1.0$\,\AA\ for H and $a_s=2.3$\,\AA\ for Fe) and we impose
a dimerization on the atomic positions of the chain, according to the
rule:

\begin{equation}
  r_n=na_s+(-1)^n \delta.
  \label{eq:dimerize}
\end{equation}
By varying the dimerization parameter $\delta$ we get the results
shown in Fig.~\ref{fig:dimerdft}, with the top panel again devoted to
H and the bottom panel to Fe, as in Fig.~\ref{fig:Edft}. Note that,
for better comparison, we have displaced vertically the energies for
each magnetization by $E_{\delta=0}(\<S_z\>)$, labeled as
$E_0(\<S_z\>)$. As we can see, in the case of zero magnetization, the
H energy presents a minimum at a dimerization parameter $\delta
\approx 0.21$\,\AA, thus confirming our conjecture: the film will
reconstruct in this case, if the substrate potential is not too
strong. On the other hand, for half and full magnetization we can see
that the energy tends to a minimum for zero dimerization, showing
that, at least, this reconstruction scheme does not reduce the total
energy. Thus, we are allowed to conjecture, based on the presented
data, that this system will show different structures for zero and for
high magnetic fields.

This phenomenon is even more salient for Fe, as we can see in the
bottom panel of \ref{fig:dimerdft}. The equilibrium value of the
dimerization parameter is strongly dependent on the
magnetization. Thus, we are led to conjecture that the imposition of
a strong external magnetic field may induce structural changes.


\section{Physical picture}
\label{sec:picture}

The combination of {\em ab initio} calculations and statistical
mechanics provides a unified physical picture of the complex physical
behavior of absorbed nanowires in presence of external magnetic fields.

The simplest scenario corresponds to hydrogen atoms on an inert and
rigid substrate, as shown both using the IFK model and DFT
calculations. In absence of an external field and for weak substrate
potentials, the chain will reconstruct while presenting an
antiferromagnetic structure. This reconstruction can be avoided by
three different routes: increasing the temperature, increasing the
external magnetic field or diminishing the ratio between the film and
the substrate potentials. The antiferromagnetic to paramagnetic
transition present the usual features associated to a long-range Ising
model.

We should ask about the ranges for the temperature and the magnetic
fields for which the phenomena discussed in this article will take
place.  In Sec.~\ref{sec:ifk} we have employed numerical values for
the physical parameters chosen to resemble the effective potentials in
the H chain.  In that case, we can see that the critical temperature
$T_c\sim 0.1$\,eV $\sim 1200$\,K, and the magnetic field $H\sim 1$\,eV
$\sim 2\cdot 10^4$\,T, which is simply too large for any practical
purposes.  In general terms, if the film is composed of atoms or
molecules with a large magnetic moment, the necessary magnetic field
will be reduced by the same factor.  Yet, the complexity of the energy
curves in multielectronic atoms can play in our favor, as it may be
the case of Fe.  As we see in Figs.~\ref{fig:Edft} (bottom), the
energy curve for zero magnetization shows the lowest values of the
energies for $a_s$ in a range from $2.8$ to $3.5$\,\AA.  Choosing an
appropriate $a_s$, the energy difference between the curves
corresponding to different magnetization levels can be made
arbitrarily low, thus allowing a small external magnetic field to
provide the necessary difference to induce a phase transition.

The nanowire of Fe atoms deserves further attention. The potential
energy curves shown in the bottom panels of Figs. \ref{fig:Edft} and
\ref{fig:dimerdft} suggest that the IFK model should be extended in
order to provide a full physical explanation of this case. The
straightforward procedure would be to fit the $V_F$ and $V_{AF}$
potentials to the numerical data obtained from DFT, but it is easy to
understand that this will not be enough. The intrincate behavior of
iron atoms can not be accounted for using classical Ising spins
($S_z=\pm1$). A classical statistical model would require, at least,
the use of or Heisenberg spins \cite{Baxter.82}.

Our calculations have shown that the general mechanism provided in
this article can work in real materials. Of course, these calculations
are only a proof-of-principle, using simple geometries. Indeed, it is
natural for atomic chains to dimerize due to Peierls instability
\cite{Peierls}, although the dimerization in our case has a
different origin. Further calculations, using more realistic
materials, are still needed in order to make any experimental proposal
to observe the predicted reconstruction effects. We should remark
that our results are in line with those of previous works
\cite{Tung,Zarechnaya}, which provide some theoretical evidence of
{\em magnetic crossovers} in transition atoms, as the preferred
magnetization varies as a function of the atomic distances.  In
some cases, the chains are more stable when displayed along a
zig-zag geometry \cite{Tung}.

An interesting experimental route may be provided by the use of
ultracold atoms in optical lattices, since most parameters can be
easily engineered, and thus the different transitions can be observed
just tuning the intensities of the laser beams
\cite{GarciaMata.07,Lewenstein.12}.


\section{Conclusions and Further Work}

We have put forward the following question: can nanowires reconstruct
differently in the presence of external magnetic (or electric) fields?
After our calculations, we can conjecture that this can indeed be the
case. We have performed illustrative {\em ab initio} calculations
using DFT, showing that this possibility exists for two types of
atoms: hydrogen and iron.

Furthermore, we have proposed a statistical mechanical model, which is
an Ising-like extension of the Frenkel-Kontorova model, the IFK model,
in which film atoms interact differently when their spin variables are
the same or opposite. We have extracted some salient physical
consequences in the 1D case, using reasonable forms for the film
potentials and a sinusoidal form for the interaction with the
substrate, showing a rich behavior with an
antiferromagnetic-paramagnetic second-order phase transition at a
finite value of the temperature. It is relevant to discuss how an
Ising-like model can give rise to a phase transition at finite
temperature in 1D, since they are forbidden for short-ranged Ising
models due to entropic considerations \cite{Huang}. The reason is as
follows: we may integrate out the spatial degrees of freedom, giving
rise to an effective Ising model for the spins presenting long-range
interactions. Indeed, the critical exponents that we have found allow
us to conjecture that, indeed, our model behaves as a long-range Ising
model in 1D.

The mechanism described in this paper bears some similarity with
colossal magnetoresistance (CMR), where metallic ferromagnetic regions
co-exists with insulating antiferromagnetic ones, due to the presence
of quenched disorder \cite{Dagotto.05}. An external magnetic field
will favor the ferromagnetic regions, thus allowing them to reach the
percolation threshold and decrease the effects the disorder and
the resistance dramatically. 

It is likely that, as we increase the magnetic field, the
antiferromagnetic configuration will not become directly unstable, but
metastable. In other terms: the transition may be of first order.
This implies that, as one cycles over a range of magnetic fields, we
will obtain a hysteresis cycle.

Throughout this article we have used a magnetic field to force the
change in reconstruction. In principle, electric fields can also be
used, in the case of film atoms or molecules with a permanent electric
dipole.

In order to proceed with this line of research, there are several
complementary routes. First of all, it will be very interesting to
consider how some characteristic features of the FK model extend to
the IFK case, such as the commensurate-incommensurate transition or
the presence of defects (e.g. kinks). Moreover, it is worth to develop
further the statistical mechanics of the IFK, both in 1D and 2D, where
the physical properties should be richer and reconstruction would be
more experimentally feasible. In order to obtain experimental
confirmation of our results, the choice of the correct materials is of
paramount importance. We can obtain some guidance from numerical
simulations combining DFT and statistical mechanical tools in order to
select those which will present a critical magnetic field within the
experimental range. In this case, more complicated potential curves
will be required, as it is shown by the DFT results for Fe, which may
be correctly described using e.g. Heisenberg spins. After some
suitable materials have been chosen and characterized, we intend to
make a concrete experimental proposal.

\begin{acknowledgments}
We would like to acknowledge Elka Korutcheva, Julio Fernández, Rodolfo
Cuerno and Pushpa Raghani for very useful discussions. E.M.F. thanks
the RyC contract (ref. RYC-2014-15261) of the Spanish Ministerio de
Economía, Industria y Competitividad. We acknowledge funding from
Spanish Government through grants PID2019-105182GB-I00 and
PGC2018-094763-B-I00.

\end{acknowledgments}


\end{document}